\documentclass[a4paper]{article}
\usepackage[pdftex]{graphicx} 
\usepackage[english]{babel}
\usepackage[latin1]{inputenc} 
\usepackage[bf,small,nooneline]{caption}
\usepackage[round]{natbib}

\newcommand{\dfrac}[2]{\frac{\displaystyle#1}{\displaystyle#2}}
\newcommand{\pder}[2]{\frac{\displaystyle\partial#1}{\displaystyle\partial#2}}

\title{Handling vacuum regions in a hybrid plasma solver}

\begin{document}
\date{November 10, 2012}
\author{M.\ Holmstr{\"o}m\thanks{Swedish Institute of Space Physics, PO~Box~812, SE-98128~Kiruna, Sweden. (\texttt{matsh@irf.se})}}
\maketitle

\begin{abstract}
In a hybrid plasma solver (particle ions, fluid mass-less electrons) 
regions of vacuum, or very low charge density, can cause problems since 
the evaluation of the electric field involves division by 
charge density. 
This causes large electric fields in low density regions that can lead 
to numerical instabilities. 
Here we propose a self consistent handling of vacuum regions 
for hybrid solvers. 

Vacuum regions can be considered having infinite resistivity, 
and in this limit Faraday's law approaches a magnetic diffusion equation. 
We describe an algorithm that solves such a diffusion equation 
in regions with charge density below a threshold value. 
We also present an implementation of this algorithm in a hybrid plasma solver, 
and an application to the interaction between the Moon and the solar wind. 

We also discuss the implementation of hyperresistivity 
for smoothing the electric field in a PIC solver. 
\end{abstract}

\section{The hybrid equations}
In the hybrid approximation, ions are treated as particles, 
and electrons as a massless fluid. 
In what follows we use SI units. 
The trajectory of an ion, $\mathbf{r}(t)$ and $\mathbf{v}(t)$, 
with charge $q$ and mass $m$, is computed from the Lorentz force, 
\[
  \dfrac{d\mathbf{r}}{dt} = \mathbf{v}, \quad
  \dfrac{d\mathbf{v}}{dt} = \dfrac{q}{m} \left( 
    \mathbf{E}+\mathbf{v}\times\mathbf{B} \right), 
\]
where $\mathbf{E}=\mathbf{E}(\mathbf{r},t)$ is the electric field, 
and $\mathbf{B}=\mathbf{B}(\mathbf{r},t)$ is the magnetic field.  
From now on we do not write out the dependence on $\mathbf{r}$ and $t$. 
The electric field is given by 
\begin{equation}
  \mathbf{E} = \dfrac{1}{\rho_I} \left( -\mathbf{J}_I\times\mathbf{B} 
  +\mu_0^{-1}\left(\nabla\times\mathbf{B}\right) \times \mathbf{B} 
   - \nabla p_e \right) + \frac{\eta}{\mu_0} \nabla\times\mathbf{B}, 
\label{eq:E}
\end{equation}
where $\rho_I$ is the ion charge density, 
$\mathbf{J}_I$ is the ion current density, 
$p_e$ is the electron pressure, 
$\eta$ is the resistivity, and
$\mu_0=4\pi\cdot10^{-7}$ is the magnetic constant. Then 
Faraday's law is used to advance the magnetic field in time, 
        \begin{equation}
          \pder{\mathbf{B}}{t} = -\nabla\times\mathbf{E}. \label{eq:F}
        \end{equation}
Note that the unknowns are the position and velocity of the ions, and 
the magnetic field on a grid, \emph{not} the electric field, since it 
can always be computed from~(\ref{eq:E}). 
Further details on the hybrid model used here, and the discretization, 
can be found in~\citet{Enumath09,Astronum10}. 

\section{Vacuum regions}
In regions of low ion charge density, $\rho_I$, the hybrid method can have 
numerical problems.  We see from~(\ref{eq:E}) that the electric field 
computation involves a division by $\rho_I$. 
In what follows we will not write out the $I$ subscript, i.e.\ $\rho=\rho_I$. 
Thus, in low density regions we will have large electric fields, 
and in the limit of zero charge density, the electric field magnitude 
will tend to infinity.  
This can lead to numerical instabilities, due to large gradients 
in the electric field, and due to large accelerations of ions. 
The solution quickly becomes unstable. 

We can either have a region where $\rho=0$ physically, like 
inside a resistive obstacle, or due to the statistical nature of
particle in cell solvers, there can be regions where we simply have 
no macro particles, or very few. 
From now on we denote all the different cases of 
low density regions as \emph{vacuum regions}. 

Many ad hoc solutions have been proposed to handle vacuum regions. 
One way is to set a minimum allowed ion charge density, e.g., 
if $\rho$ is below a threshold value in a cell, $\rho$ is set to 
that value for the cell. 
A similar solution is to set a maximum allowed value for the 
electric field.  
One can also introduce new ion sources that were not part of 
the original problem, to keep $\rho$ large enough in all cells. 
If we have an absorbing obstacle we can instead of removing 
absorbed ions reduce their weight gradually over time toward zero. 
None of these solutions solve the original problem. 
In the case of threshold values we do not get the solution to 
the hybrid equations, and in the case of artificial sources 
or losses, we are solving the equations self consistently, but 
we are solving a different physical problem. 

However, a self consistent, physically correct way of handling the 
problem of vacuum regions was proposed a long time ago. 
\citet{Hewett} noted that vacuum regions can be viewed as having 
infinite resistivity, and an algorithm can be devised where 
the resistivity in~(\ref{eq:E}) is set to a large value in regions 
of low density.  \citet{Harned} also used a similar idea and solved 
a Laplace equation for the electric field in vacuum regions. 
However, solving Laplace equation over a complicated region that is 
also changing over time is not an easy task.  
Especially if one wants to do it on a parallel computer, since 
solving Laplace equation on a grid involves solving 
a system of linear equations. 

\section{Solving a diffusion equation in vacuum regions}
Building on the idea of \citet{Hewett} of having a large resistivity 
in low density regions, let us see what a large resistivity implies. 

Let the resistivity be variable in space, and time, 
$\eta=\eta(\mathbf{r},t)$. 
If we assume that the resistive term dominate in the 
expression~(\ref{eq:E}) for the electric field, then 
Faraday's law~(\ref{eq:F}) becomes
\[
  \pder{\mathbf{B}}{t} = -\nabla\times 
    \left( \frac{\eta}{\mu_0} \nabla\times\mathbf{B}  \right). 
\]
For a constant resistivity we have that 
\begin{equation}
  \pder{\mathbf{B}}{t} = \frac{\eta}{\mu_0} \nabla^2 \mathbf{B}. 
    \label{eq:heat}
\end{equation}
This is a diffusion equation for the magnetic field, and the steady 
state solution will be a solution to the Laplace equation 
$\nabla^2 \mathbf{B}=0$.

\section{Time step limits for the diffusion equation} \label{sec:dt}
The diffusion equation for the magnetic field~(\ref{eq:heat}) is similar to 
the heat equation and gives a time step limit for stability of 
\[
  \Delta t < \frac{\mu_0\Delta x^2}{2\eta}, 
\]
where $\Delta t$ is the time step for an explicit time integrator. 
For large resistivities this 
will set the limit for the allowed length of the time step, e.g., 
if we increase $\eta$ by 10, 
$\Delta t$ needs to be decreased by 10 for stability.

\section{Algorithm}
The details of the algorithm for handling low density regions is as follows. 
We assume that we have a resistivity, $\eta=\eta(\mathbf{r},t)$, 
that is defined in all regions of the solution domain. 
A cell is denoted a \emph{vacuum cell} if $\rho<\rho_v$ for the cell. 
The electric field in a vacuum cell is computed by~(\ref{eq:E}), 
with $1/\rho=0$, i.e.\ only the resistive term is non-zero. 
If the vacuum cell is outside any obstacle, then we set the resistivity to 
a vacuum resistivity, $\eta=\eta_v$. 
If the vacuum cell is in an obstacle, we keep the original resistivity. 

The advantage of such an approach is that vacuum regions and 
regions of a resistive object can be handled by the original solver. 
The time advance of Faraday's law~(\ref{eq:F}) is the same for all cells. 
We only compute the electric field differently in different cells.

\section{Hyperresistivity}
The leapfrog time stepping scheme that we use to discretize 
Faraday's law in time has no numerical diffusivity. 
To stabilize the computations one may have to smooth the solution. 
Here we choose to use a hyperresistivity term in the 
electric field computation. 

Hyperresistivity of different orders were introduced 
by~\citet{Maron} for an MHD solver. 
Here we introduce a hyperresistivity, $\eta_h$, and the 
expression for the electric field~(\ref{eq:E}) becomes
\begin{equation}
  \mathbf{E} = \dfrac{1}{\rho_I} \left( -\mathbf{J}_I\times\mathbf{B} 
  + \mathbf{J} \times \mathbf{B} 
   - \nabla p_e \right) + \eta \mathbf{J} 
   - \eta_h \nabla^2 \mathbf{J}, 
\label{eq:Ehyp}
\end{equation}
where the current, $\mathbf{J}=\mu_0^{-1}\nabla\times\mathbf{B}$. 
The advantage of hyperresistivity compared to smoothing, that is 
often implemented in hybrid solvers, is that the higher order 
derivatives of the Laplacian diffuse high frequency structures, 
while preserving lower frequency ones~\citep{Maron}. 
Also, the addition of hyperresistivity can often increase the 
maximum stable time step. 
In our implementation, the Laplacian 
\[ 
  \nabla^2 \mathbf{J} = \left( \pder{^2J_x}{x^2} 
  \pder{^2J_y}{y^2} \pder{^2J_z}{z^2} \right)^T
\]
is discretized using standard second order finite difference stencils. 
The hyperresistive term is present in all parts of the domain, 
also in the vacuum regions, and in the examples that follows, 
we have used a value of $\eta_h=5\cdot 10^{14}$.

\section{An application: The Lunar plasma wake}
A good example where vacuum and low density regions occur is 
the plasma wake behind the Moon, formed by the absorption on 
the dayside of the impinging solar wind plasma.  
More details on a hybrid model of the interaction between 
the Moon and the solar wind can be found in~\citet{EPS}. 
The simulation parameters used here are similar to those in that work, 
with an interplanetary magnetic field (IMF) at a $45^{\circ}$ angle 
to the solar wind flow. 

\begin{figure}[p]
\begin{center}
  \includegraphics[width=0.47\columnwidth]{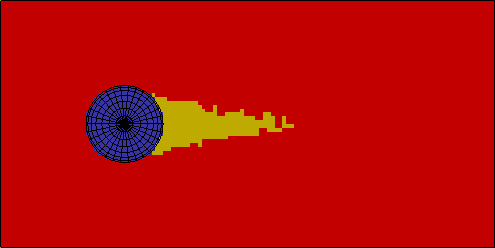}
\end{center}
\caption{
Resistivity in different regions of the simulation domain at $t=30$~s. 
A cut in the plane of the IMF. 
The solar wind flows in from the left. 
The blue region is the interior of the Moon with $\eta=10^7$~$\Omega$m. 
Red shows solar wind plasma with $\eta=0$, and in yellow is 
the wake region with $\eta_v=10^8$, where $\rho<\rho_v=0.0001$. 
} \label{fig:res}

\begin{center}
  \includegraphics[width=0.65\columnwidth]{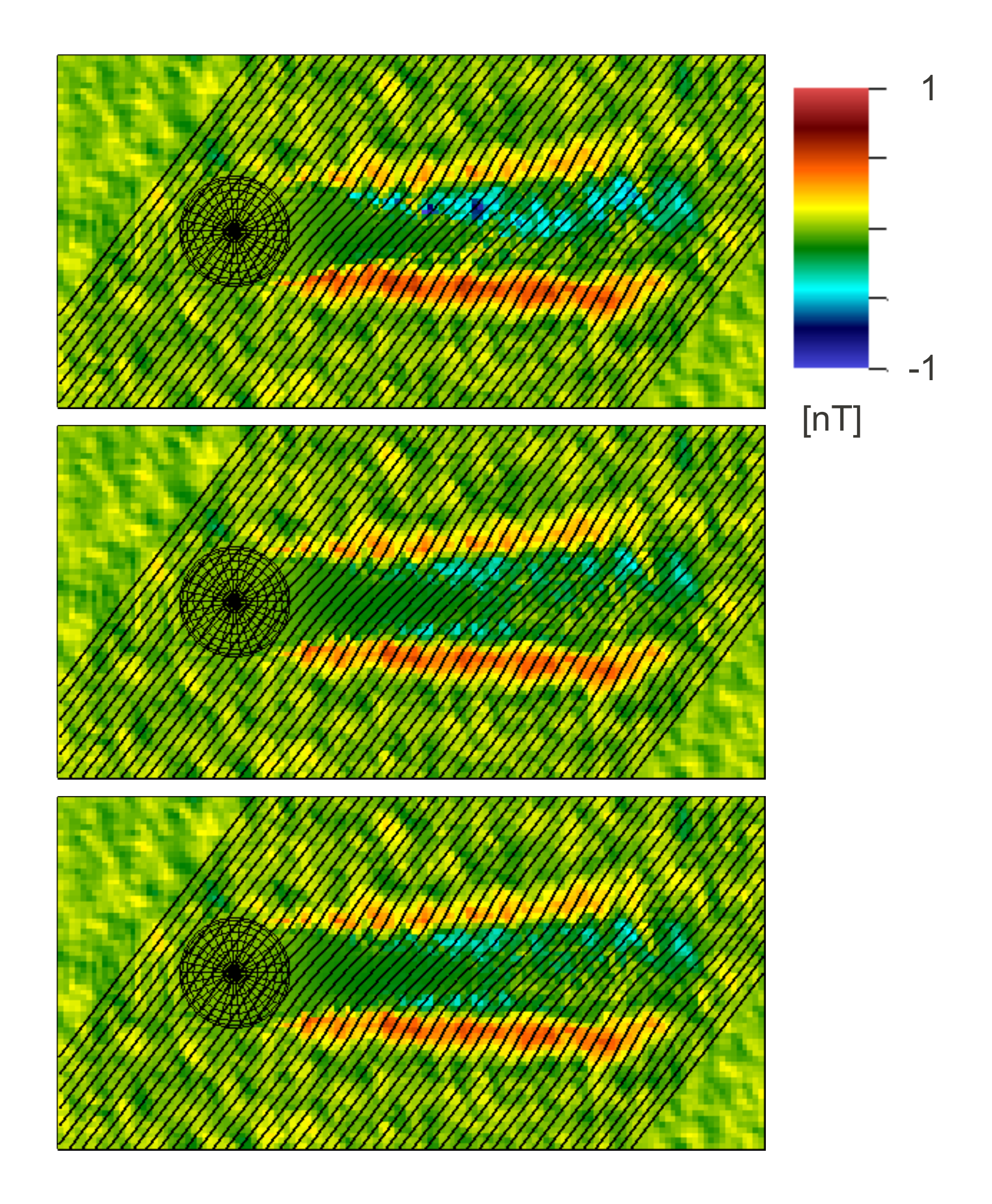}
\end{center}
\caption{
The effect of different minimum charge densities, $\rho_v$. 
Shown is the component of the magnetic field that is perpendicular 
to the plane containing the IMF. 
The minimum charge density, $\rho_v$, decrease from top to bottom, 
with values of 0.01, 0.001, and 0.0001. 
The cuts are planes that contain the IMF, and the black lines are 
the magnetic field lines. 
The solar wind flows in from the left. 
} \label{fig:cmin} 
\end{figure}

The different resistivity regions at $t=30$~s are shown in Fig.~\ref{fig:res}. 
Here $\eta$ is defined to be 0, except for the interior of the Moon 
(a sphere of radius 1730~km) where $\eta=10^7$~$\Omega$m. 
Cells with a relative ion charge density less than 0.0001, 
that are outside the obstacle (the Moon), was 
considered vacuum cell ($\rho_v=0.0001$), and there the 
resistivity was set to a vacuum resistivity $\eta_v=10^8$. 
We clearly see that the vacuum region is irregular. 
It also changes with time. 

By following the above approach to handle vacuum regions, 
we have introduced two new numerical parameters. 
A threshold ion charge density, $\rho_v$, below which we 
consider a cell to be a vacuum cell; and a vacuum resistivity, 
$\eta_v$, that we set the resistivity to in such cells. 
Ideally, $\eta_v$ should be as large as possible and $\rho_v$ 
should be as small as possible, for the solution to approach 
the solution to the original hybrid equations. 
What limits the value of $\eta_v$ is the time step limit 
discussed in Section~\ref{sec:dt}. The computational time 
will increase in proportion to $\eta_v$ for an explicit time integrator. 
The effect of different minimum density, $\rho_v$, 
are shown in Fig.~\ref{fig:cmin}. 
We see that $\rho_v$ has to be small enough to get a smooth transition 
between vacuum and non-vacuum regions. 
The solution converges as $\rho_v$ is decreased, 
as can be seen especially in the central wake region. 
We can note that the relative $\rho_v$ is as small as 
one part in 10000 for the converged solution. 
This can be contrasted with the approach of just setting a minimum 
charge density.  Then one typically has to choose a value of the 
relative density on the order of a few \%, otherwise numerical 
instabilities will develop. 
In contrast, the magnetic diffusion equation has to be applied 
only in cells of very low density. 
The method is also stable in the presence of large gradients in 
the resistivity.  The example shown in 
Fig.~\ref{fig:res} and~\ref{fig:cmin} has 
a step change in resistivity, from 0 to $10^7$ at the Lunar surface. 

\section{Summary and conclusions}
We have shown that vacuum regions in a plasma particle in cell solver can be 
handled by setting a high resistivity in those regions. 
This leads to the solution of a magnetic diffusion equation 
in such regions.  A minimum charge density parameter decide 
what cells are vacuum, where the magnetic diffusion equation 
should be advanced in time.  It is also possible to include 
arbitrary resistive obstacles.  
The algorithm was exemplified by modeling the plasma interaction 
between the solar wind and the Moon. 

The advantage of the method is that it handles vacuum regions and 
resistive obstacles in a self consistent manner. 
Also, the magnetic diffusion equation only has to be applied 
in very low density cells, e.g., a relative density of 0.0001, 
compared to a reference density.  The method seems stable to 
discontinuities in the resistivity. 

A disadvantage is that the time advance of the magnetic diffusion equation 
requires a small time step.  This is however needed anyway for some 
problems, e.g., if we model the Moon and want to include 
crustal magnetic anomalies. 
A solution for future study would be to use an implicit time integrator 
to advance Faraday's law.

\subsection*{Acknowledgments} 
This research was conducted using resources provided by the Swedish National 
Infrastructure for Computing (SNIC) at the High Performance Computing Center 
North (HPC2N), Ume\aa\ University, Sweden.
The software used in this work was in part developed by the 
DOE-supported ASC / Alliance Center for Astrophysical 
Thermonuclear Flashes at the University of Chicago.

\end{document}